\documentclass[twocolumn,english,british]{revtex4}
\usepackage[LGR,T1]{fontenc}
\usepackage[latin9]{inputenc}
\usepackage[letterpaper]{geometry}
\usepackage{amstext}
\usepackage{graphicx}
\usepackage{subscript}

\makeatletter

\DeclareRobustCommand{\greektext}{%
  \fontencoding{LGR}\selectfont\def\encodingdefault{LGR}}
\DeclareRobustCommand{\textgreek}[1]{\leavevmode{\greektext #1}}
\DeclareFontEncoding{LGR}{}{}
\DeclareTextSymbol{\~}{LGR}{126}

\@ifundefined{textcolor}{}
{%
 \definecolor{BLACK}{gray}{0}
 \definecolor{WHITE}{gray}{1}
 \definecolor{RED}{rgb}{1,0,0}
 \definecolor{GREEN}{rgb}{0,1,0}
 \definecolor{BLUE}{rgb}{0,0,1}
 \definecolor{CYAN}{cmyk}{1,0,0,0}
 \definecolor{MAGENTA}{cmyk}{0,1,0,0}
 \definecolor{YELLOW}{cmyk}{0,0,1,0}
}

\makeatother

\usepackage{babel}
\begin{document}

\title{Remote RF excitation for a small-bore MR imager at 15.2 T\thanks{This work was partially presented at the The Joint Meeting of the
International Society for Magnetic Resonance in Medicine and European
Society of Magnetic Resonance in Medicine and Biology, Paris, France,
16-21 June 2017.} }

\author{\textbf{F. Vazquez\textsuperscript{1}, S. E. Solis-Najera\textsuperscript{1},
J. Lazovic\textsuperscript{2}, L. M. Zopf\textsuperscript{2}, R.
Martin\textsuperscript{\selectlanguage{english}%
1\selectlanguage{british}%
}, L. Medina\textsuperscript{\selectlanguage{english}%
1\selectlanguage{british}%
}, O. Marrufo\textsuperscript{\selectlanguage{english}%
3\selectlanguage{british}%
}, A. O. Rodriguez\textsuperscript{4}}\thanks{Corresponding author: Alfredo O. Rodriguez, arog@xanum.uam.mx}\medskip{}
}

\address{$^{\text{1}}$Physics Department, Faculty of Sciences, Universidad
Nacional Autonoma de Mexico, Mexico City 04510, Mexico. \textsuperscript{\selectlanguage{english}%
2\selectlanguage{british}%
}Campus Science Support Facilities GmbH, Austria. \textsuperscript{\selectlanguage{english}%
3\selectlanguage{british}%
}Department of Neuroimage, INNN MVS, Mexico City, Mexico. \textsuperscript{4}Department
of Electrical Engineering, Universidad Autonoma Metropolitana Iztapalapa,
Mexico City 09340. Mexico. }

\maketitle
\medskip{}

\textbf{Abstract}. The travelling-wave MRI approach is an alternative
to overcome the B1 inhomogeneity at UHF MRI for human applications.
More recently, this concept has been also used with animal MR imagers.
We used a parallel-plate waveguide and a bio-inspired surface coil
to generate mouse images at 15.2T. Ex vivo mouse images were acquired
without high dielectric materials to conduct the signal at the right
frequency. These results are in very good concordance with results
obtained at 3T and 9.4T. These results show that travelling wave MRI
with high SNR can be performed with a simple waveguide.

\section{Introduction }

The limited sensitivity of magnetic resonance imaging (MRI) originates
a poor spatial and spectral resolutions. These limitations can be
overcome with high field MR imagers which enable the research of human
organ structure, function and chemistry under unprecedented conditions
{[}1-4{]}. The B1 inhomogeneity inside the sample to be imaged at
ultra high field MRI for human applications is a major drawback. This
is an vital aspect because the RF field wavelight approaches the size
of the sample. Another important matter is the power deposition in
the human tissue/organ too. The travelling-wave MRI (twMRI) approach
is an alternative to overcome the B1 inhomogeneity at ultra high field
MRI for human applications {[}5-6{]}. 

A very limited number of laboratories have access to UHF MR imager
sfor human imaging across the world. However, preclinical imagers
are a commodity and many research groups have access to perform research
in different areas of human and animal model diseases. The installed
preclinical MR imager base has been designed to accommodate high field
magnets raging from 4.7 T upto 21 T, altough there is a number of
higher field magnet in development {[}7{]}. The large experience gained
over the past decades on the pros and cons of preclinical UHF MRI
may prove useful for the development of UHF MRI for humans. Preclinical
systems have been developed for UHF MRI to study various diseases
using animal models {[}8{]}. More recently, this concept of twMRI
has been also used with animal MRI systems {[}9-11{]}. 

This approach offers the advantages of keeping the high volatges away
from the patient and nuclei are more homogenoulsy excited. Under standard
circunstances only one transceiver coil is used for the entire MRI
experiment. This might be an important limitation because of the low
SNR produced, then, high fields are mandatory to partially overcome
this disavantage. The critical frequency of the waveguide to transmit
the energy depends on the dimensions of the waveguide itslef. This
is a major problem, because the dimensions may be larger than the
magnet bores commonly found in both human and small-bore MR imagers.
Dielectric paddings with high permittivity have been successfuly used
tu run twMRI experiments {[}11{]}.

We have previously demonstrated that the PPWG can transmit the RF
signal at 3 T {[}6{]} and 9.4 T {[}10{]} without dielectric materials,
and that the waveguide dimenions play no role to run twMRI experiments.
These encouraging results motivated to put the PPWG to the test at
higher resonant frequencies. We investigated the use of a parallel-plate
waveguide (PPWG) and a bio-inspired surface coil to generate mouse
images with a preclinical MR imager at 15.2 T. We demonstrate that
the PPWG can transmit all frequencies at 650 MHz (proton frequency
at 15.2 T) because its critical frequency is zero.

\section{Waveguide and RF coil }

We built a PPWG To test the viability of this remote RF excitation
approach at 15.2 T. This waveguide offers the advantage that all frequencies
can propagate because its critical frequency is zero for the TE principal
mode. The PPWG prototype was built using 2 aluminium strips (4 cm
wide and 6 lm thickness) were mounted on an acrylic tube with a 3
cm diameter and 1.2 m long. 

The mechanical properties of the first prototype allow us to have
constant cross-section for this particular length (Fig. 1). The prototype
was used together with an RF surface coil located at one end of the
waveguide and an aluminium blocker was at the opposite end. The coil
prototype consisted of 6 circular petals (0.45 cm diameter) and a
total radius of 1 cm, then it was matched and tuned to 50 \textgreek{W}
and 650 MHz. Fig. 1 shows an illustration of the surface coil and
the experimental setup. 

\medskip{}
\begin{minipage}[t]{1\columnwidth}%
\begin{center}
\includegraphics[scale=0.3]{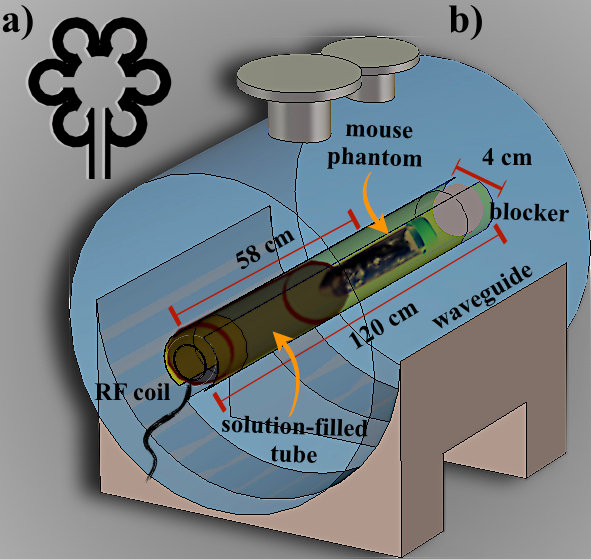}
\par\end{center}

\begin{center}
\textbf{\small{}Figure 1. a). Surface coil for transmission of the
RF signal. b). Experimental setup for all twMRI experiments. }
\par\end{center}{\small \par}%
\end{minipage}\medskip{}

Coil performance was measured via the quality factor, \emph{Q}, and
the coil noise figure. To test both the feasibility of remore RF excitation
approach and performance of the MR scanner, we performed exvivo imaging
experiments with a formaldehyde-fixed mouse phantom. 

To assure the transmission of the energy inside the PPWG {[}12{]},
saline-solution filled tubes and the mouse phantom were inserted inside
the waveguide, see Fig. 1.

\subsection{Waveguide testing}

An electromagnetic semi-anechoic chamber was use to test the viability
of the PPWG. The chamber is 3m x 3m and it can be operated from 30
MHz to 1 GHz with a normalized attenuation of $\pm$4 dB and, from
1.1 GHz to 18 GHz with a normalized attenuation of $\pm$3 dB. Fig.
shows an illustration of the testing setup using both the waveguide
and the RF surface coil. The PPWG was air-filled and the surface coil
was linealrly polarized for all testing experiments. 

\medskip{}

\begin{minipage}[t]{1\columnwidth}%
\begin{center}
\includegraphics[scale=0.2]{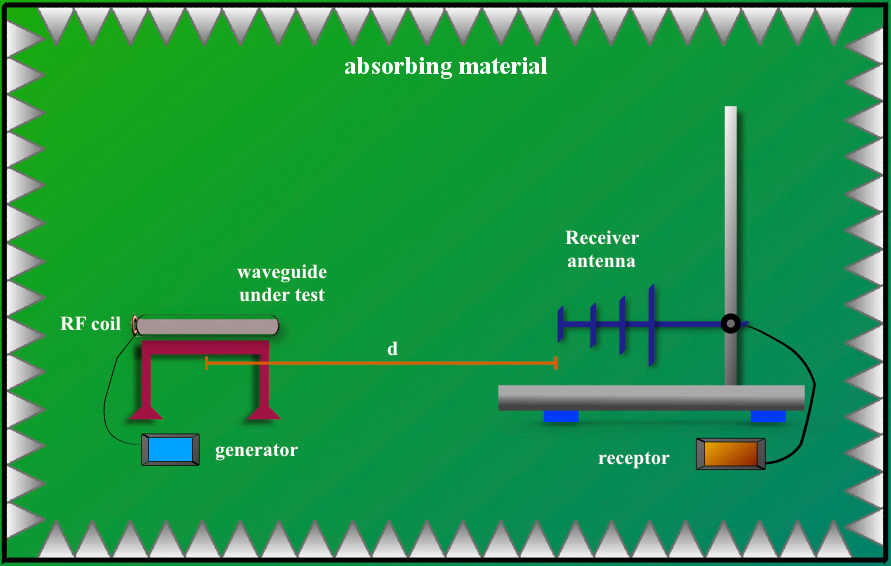}
\par\end{center}

\begin{center}
\textbf{\small{}Figure 2. Setup of the PPWG to test its viability
in a semi-anechoic chamber. The PPWG was located on a turntable 2
m away from a bilog receiver antenna, and the transmission RF coil
was at the opposite end of the waveguide.}
\par\end{center}{\small \par}%
\end{minipage}

\medskip{}

\subsection{Imaging experiments}

A mouse (40g) was used for ex vivo imaging experiments. The mouse
was positioned at the magnet isocentre and was 56 cm away form the
coil, as shown in Figure 1.b). We used a GE sequence (FLASH) and the
following acquisition parameters: TR/TE=100/1.6ms, FA=250, FOV=18x18mm2,
matrix size=256x256, thickness=1mm, NEX=2. All experiments were run
in a 15.2T/11cm MR imager (BioSpec, Bruker Co, Ettlingen, Germany).

\section{Results and Discussion }

The coil quality factors were approximately: \emph{Q\textsubscript{\emph{u}}}/\emph{Q\textsubscript{\emph{l}}
}= 21/13, and its noise figure (NF) was 10.5. This bench test results
and the NF show a good performance of the bio-inspired coil. Radiation
patterns were acquired for two differences distance and orientations,
shown in Fig. 3. As expected, because the bioinspired surface coil
has a symmetrical configuration, so the radiation patttern is necessary
symmetrical. 

\medskip{}
\begin{minipage}[t]{1\columnwidth}%
\begin{center}
\includegraphics[scale=0.33]{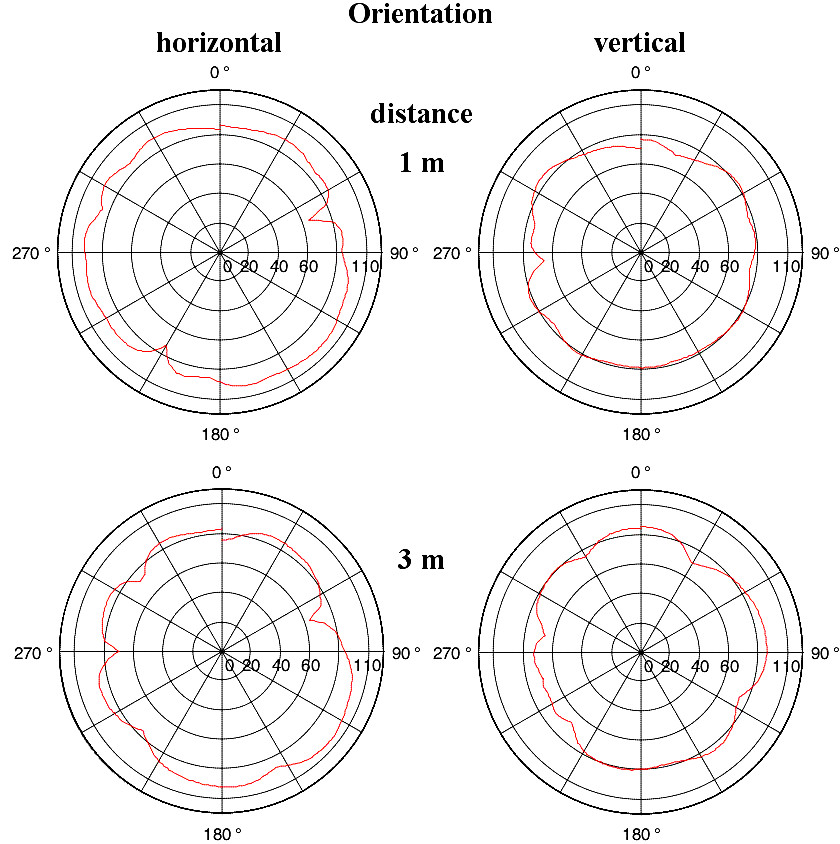}
\par\end{center}

\begin{center}
\textbf{Figure 3. Polar diagrams of amplitude patterns for the vertical
and horizontal planes for the bioinspired surface coil and the PPWG.}
\par\end{center}%
\end{minipage}

\medskip{}

Ex vivo mouse images were acquired with the PPWG and our coil prototype,
see Fig. 4. Standard gradient sequences with a low flip angle were
used to acquire ex vivo mouse images. These images showed very good
image quality with clear delineation of anatomical structures, and
compatibility with standard pulse sequences.

\medskip{}
\begin{minipage}[t]{1\columnwidth}%
\begin{center}
\includegraphics[scale=0.8]{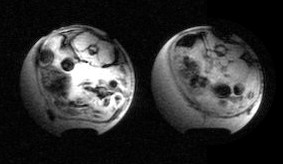}
\par\end{center}

\begin{center}
\textbf{Figure . Ex vivo mouse images acquired with the parallel-plate
waveguide and bio-inspired coil.}
\par\end{center}%
\end{minipage}\medskip{}

There was no need to use high dielectric materials to conduct the
signal at the right frequency {[}2{]}. These results are in very good
concordance with results obtained at 3 T {[}4{]} and 9.4 T {[}5{]}.
From these results, we can conclude that the MR signal can be transmitted
regardless the resonant frequency and the magnet bore size, to acquire
mouse images at ultra high field. In vivo experiments can be conducted
under more comfortable conditions for the animal model. These results
shows that travelling wave MRI with high SNR can be performed with
a simple waveguide only.

\section{Conclusions}

We have experimentally demonstrated that the use of PPWGs can produce
good SNR images for relatively large fields of view via the travelling-wave
approach. We have shown that the waveguide approach can also be used
with magnetic field intensities larger than 7T and samll-bore MRI
systems. Overcoming the limitation of a cutoff frequency, as demonstrated
in this work, provides further freedom of implementation in a variety
of geometries and for multinuclear operation, or for measuring several
or extended samples. Further investigation is required to explain
the physical mechanisms involved in the RF signal with a dielectric
non homogenous object inside a waveguide and its implications on image
quality. A natural step ahead is to extend this approach to acquire
images of the entire human body. These results pave the way to a further
implementation of the travelling-wave approach in other applications
using lower magnetic field intensities and multinuclear experiments
with bores of different dimensions.

\section{Acknowledgments}

We thank CONACYT Mexico (grant 112092) and, PAPIIT-UNAM (grant IT
102116), and the Preclinical Imaging Facility at Vienna Biocenter
Core Facilities (VBCF), Austria. 

\bigskip{}


\begin{thebibliography}{10}
\bibitem{key-1}Niendorf T, et. al. (2016) From ultrahigh to extreme
field magnetic resonance: where physics, biology and medicine meet.
29: 309-311.

\bibitem{key-2}U\u{g}urbil K. (2014) Magnetic resonance imaging at
ultrahigh fields. IEEE Trans. Biom. Eng. 61: 1364-1379.

\bibitem{key-3}Polenova T, et. al. (2016) Ultrahigh field NMR and
MRI: Science at a crossroads. Report on a jointly-funded NSF, NIH
and DOE workshop, held on November 12-13, 2015 in Bethesda, Maryland,
USA. J Mag. Reson. 266: 81-86.

\bibitem[4]{key-4}Markiewicz WD, et. al. (2015). A Decade of Experience
With the UltraWide-Bore 900-MHz NMR Magnet. IEEE Trans. Appl. Supercond.
25: 1-5.

\bibitem[5]{key-5}Brunner DO, et. al. (2009). Travelling-wave nuclear
magnetic resonance. Nature. 457: 994-998.

\bibitem[6]{key-6}Vazquez F, et. al. (2013) Travelling wave magnetic
resonance imaging at 3 T. J. App. Phys. 114: 0649006. 

\bibitem[7]{key-7}Moser E, et. al. (2017). Ultra-High Field NMR and
MRI\textemdash The Role of Magnet Technology to Increase Sensitivity
and Specificity. Front. Phys. 5: 33.

\bibitem{key-9}Marzola P, et. al. (2003) High field MRI in preclinical
research. Eur. J. Radiol. 48: 165-170.

\bibitem[8]{key-8}Bluem P, et. al. (2015) Travelling-wave excitation
for 16.4 T small-bore MRI. IEEE MTT-S Intern. Microwave Symp: 1. 

\bibitem[9]{key-9}Tonyushkin AA, et. al. (2012) Traveling Wave MRI
at 21.1 T: Propagation below Cut-off for Ultrahigh Field Vertical
Bore System. Proc. Intl. Soc. Mag. Reson. Med. 20: 2693. 

\bibitem[10]{key-10}Vazquez F, et. al. (2016) Travelling-wave transmitted
with a simple waveguide for rodents Magnetic Resonance Imaging at
9.4 T. 33rd Ann. Meet. ESMRMB 32: S31-S32.

\bibitem[11]{key-11}Bluemink JJ, et. al. (2016) Dielectric waveguides
for ultrahigh field magnetic resonance imaging. Magn. Reson. Med.,
76: 1314-1324.

\bibitem[12]{key-12}Vazquez F, et. al. (2013) Feasibility numerical
study of the travelling wave MRI at 3T. Proc. Intl. Soc. Mag. Reson.
Med. 21: 4371. \end{thebibliography}
\end{document}